\documentstyle[preprint,revtex]{aps}
\draft
\newcommand{\be}{\begin{eqnarray}}
\newcommand{\ee}{\end{eqnarray}}
\begin{document}
\preprint{UAHEP948}
\noindent
\vspace*{20mm}
\begin{title}
\centerline {Light Gluinos and $\Gamma(Z\rightarrow b\overline{b})$}
\end{title}
\author{L. Clavelli}
\begin{instit}
Department of Physics and Astronomy\\
University of Alabama\\
Tuscaloosa, Alabama 35487\\
\end{instit}
\centerline{Abstract}
\begin{abstract}
     We discuss the anomaly in the $b$ branching ratio of the
     $Z$ in the context of the light gluino scenario.\\
\end{abstract}
\narrowtext
\par
     Recently, some attention \cite{Miquel} has been given to a
possible deviation from the standard model in the $b\overline{b}$
decay rate of the $Z$. The discrepancy may be discussed in terms
of the ratio\\
\be
 R_b \equiv {{\Gamma(Z\rightarrow b{\overline b})}
 \over{\Gamma(Z\rightarrow hadrons)}} \label{eq:Rb}
\ee
where the current experimental results and theoretical expectations
are \cite{Batley,Chivukula,Langacker}
\be
  R_b^{exp} = .2192 \pm .0018             \label{eq:Rbexp}
\ee
\be
R_{b}^{th} =\left\{\begin{array}{ll} .2157\pm.0004 \quad &(for
M_t=174 GeV)\\.2165\pm.0004 \quad &(for M_t=150 GeV)
\end{array}\right\} \quad .\label{eq:Rbth}
\ee
Over the relevant range of $M_t$, $R_b^{th}$
is approximately linear \cite{Chivukula} so we may
parametrize the anomaly as
\be
   \delta R_b\equiv R_b^{exp} - R_b^{th} = \Bigl(18\pm 18 +
({{M_t-123}\over 3})\Bigr)\cdot 10^{-4}       \label{eq:delRb}
\ee
Thus the anomaly disappears at the one standard deviation level if
$M_t\approx 123$.
This however is below the lower limit of $131$ $GeV$ quoted by the
$D\emptyset$ \cite{D0} collaboration for the top mass and is
inconsistent with the top interpretation
of the $CDF$ \cite{CDF} results suggesting $M_t\approx 174$.\\
\par
     The anomaly may disappear with further statistics or may be
due to a theoretical underestimate of the b production from gluon
hadronization. On the other hand, it is attractive to examine each
discrepancy from the standard model in terms of a possible
explanation in supersymmetry ($SUSY$). In particular it has been
pointed out that an enhanced $b\overline{b}$ production at the $Z$
could be a signal for low lying stop quarks, charginos, or
neutralinos due to loop effects \cite{BF,BR,WKK}. Such explanations
require stop quarks, charginos, or neutralinos below $M_Z$. In the
standard $SUSY$ model with heavy squarks and gluinos an explanation
of the anomaly requires abandoning the conventional supergravity
(SUGRA) inspired model for $SUSY$ breaking \cite{WKK}. On the other
hand, in the light gluino scenario, the SUGRA inspired breaking
leads automatically to charginos and neutralinos in the required
mass region.  There are three scenarios associated with light gluinos
which could separately or together explain the excess bottom
production at $LEP$.  These are \hfill\break
\begin{itemize}
\item{\noindent 1.\hspace{1mm}} The top quark might be light (on the
order of $123$ $GeV$) thus reconciling the data through
\ref{eq:delRb}. This is ruled out by the $D\emptyset$ and $CDF$
experiments if the gluino is heavy but is allowed in the light
gluino case as discussed below.\\
\item{\noindent 2.\hspace{1mm}}The $Z$ branching ratio into $b$'s
could be enhanced by loop effects with virtual stop quarks and
charginos and/or charged Higgs as, for example, in $fig. 1$.
This has been discussed by several authors and might (marginally)
explain the effect if the stop quarks and charginos are light.  This
is natural in the light gluino scenario but otherwise requires
abandoning the current ideas about supergravity related $SUSY$
breaking.\\
\item{\noindent 3.\hspace{1mm}} The $Z$ could be decaying into
on-shell neutralinos or charginos which would give some preference
in their decay to bottom quarks through their higgsino components.
In the light gluino case these particles decay predominantly into
$q\overline{q}\tilde{g}$. In the heavy gluino case these particles
are expected to decay down into the lightest neutralino which will
then escape the detector leading to an appreciable missing energy.
This latter possibility is ruled out if the gauginos are below
$M_Z/2$ due to the $LEP$ experiments.  We reserve a quantitative
discussion of this possibility to a later paper.\hfill\break
\end{itemize}
\par
     It is generally recognized that there are one or more low mass
gluino windows consistent with current experimental limits \cite{UA1}.
for an update on these windows see \cite{Clavelli,Farrar}. However,
perhaps because these windows are small compared with the unbounded
region above $150$ $GeV$, most theoretical and experimental analyses
on squark and gluino mass limits assume that the gluino mass is not
in the low energy windows.  If, on the other hand, the gluino is
light (which we will take in this article to mean below $5$ $GeV$),
most of the limits quoted on the other $SUSY$ particles and the top
quark are also voided. Such a light gluino will decay into a
$q\overline{q}\tilde\gamma$ without an appreciable missing energy.
Therefore the squarks decaying into $q\tilde g$ would also not
produce enough missing energy to make the collider experiments
sensitive to them.  Similarly, in the light gluino scenario with a
$SUSY$ scale below approximately $570$ $GeV$, one of the stop quarks
is typically lighter than the top in which case the top decays
through the chain \cite{CC1}
\be
\begin{array}{rl}t \rightarrow &\tilde{t}\tilde{g}\cr
           \tilde{t} \rightarrow &\tilde{W} b\cr
 \tilde{W} \rightarrow & q{\overline q} \tilde{g}\cr\end{array}
 \label{eq:chain}
\ee
Such a decay chain involves no high $P_{\perp}$ leptons and would
make the top quark invisible to the current FNAL searches.  It
could therefore lie considerably lower than the $131$ $GeV$ lower
limit for a top with a standard model decay chain.  In the light
gluino scenario with radiative electroweak symmetry breaking as
currently understood, the running top quark mass is predicted to
be about $114$ $GeV$  \cite{LNW}. This corresponds to a physical
top quark mass of about $124$ $GeV$ \cite{CCMNW} Such a light top
quark mass could by itself resolve the $R_b$ anomaly as can be seen
in \ref{eq:delRb}. However as mentioned above there are other
features of the light gluino scenario which lead naturally to an
enhanced $R_b$. Any one of these, or all of them together, could
explain the discrepancy.\hfill\break
\par
     An enhancement in $\Gamma (Z \rightarrow b \overline{b})$
is naturally coupled to a higher apparent $\alpha_3$ from the $Z$
width.  Assuming that only the $b$ channel is enhanced leads to
the relation
\be
 \delta \alpha_3 = \pi \delta R_b {{1+\alpha_3/\pi}\over{1-R_b}}
   \label{eq:delalf3}
\ee
The $LEP$ result from the $Z$ width assuming a nominal top quark
mass of $174$ $GeV$, namely $\alpha_3(M_Z)=.125\pm .005$
\cite{Miquel} correlates well with the observed enhancement in
$R_b$ and the value $\alpha_3(M_Z)\approx .11$   \cite{CC2}
found by extrapolating from low energy data assuming a light
gluino.  In references \cite{CCFHPY,BR} an attempt was made to
explain the $\alpha_3$ discrepancy by a light gluino and squarks
below the $Z$ (but above $M_Z/2$ of course) through the
${\overline q}\tilde{q}\tilde{g}$ decays of the $Z$. In the SUGRA
model it is not possible to have $b$ squarks significantly below
the other squarks and sleptons and hence a low $SUSY$ scale with
a light gluino will not selectively enhance $R_b$ through this
mechanism.  In view of the current data on $R_b$, the universal
scalar mass must therefore be at least $80$ $GeV$.  The lightest
$SUSY$ scalar apart (possibly) from one of the stop quarks is
then the sneutrino with a mass above $62$ $GeV$.\hfill\break
\par
     In reference \cite{CCY}
it was noted that in the light gluino scenario with a universal
gaugino mass $M_{1/2}$ set to zero, there is a stringent limitation
on the possible values of $\tan(\beta)$. These come from the
theoretical expression for the chargino masses \cite{reviews}
\be
   M_{\chi_{2,1}^{\pm}} = {1\over 2} \Bigl (2M_W^2+\mu^2\pm
   \sqrt{\mu^2+ 4 M_W^2 \mu^2 + 4 M_W^4 {\cos^2{(2\beta)}}} \Bigr )
   \label{eq:mchi}
\ee
together with the experimental requirement that these masses be
greater than (or almost equal to) half the $Z$ mass.  Assuming
$\tan(\beta)>1$ as required in the model of radiative electroweak
symmetry breaking and taking the experimental lower limit on the
light Higgs boson to be $42$ $GeV$ \cite{PDG94} yields a possible
range for $\tan(\beta)$ of\hfill\break
\be
    1.5 < \tan(\beta) < 2.266           \label{eq:tanb}
\ee
\par
     In reference \cite{LNW} it was argued that, in the light gluino
scenario, the lower limit on the light Higgs is actually $60$ $GeV$
which then, at the current level of theoretical analysis, requires,
assuming also radiative electroweak breaking, that $\tan(\beta)$
be restricted to a very tiny window around the center of the above
range.  We are reluctant to attribute such precision to the current
perturbative arguments so we will, in the present article, explore
the already narrow range of \ref{eq:tanb}.  In ref.\cite{WKK} it was
found that the $\delta R_b$ discussed above could be brought to
within one standard deviation of zero with a top mass of $174\pm 15$
if the lightest chargino was below $60$ $GeV$ and $\tan(/beta)$ were
below $1.8$ (See their $fig. 2$).  Such a situation is not expected
in the usual treatment of the $MSSM$ with a supergravity inspired
$SUSY$ breaking and a heavy gluino.  On the other hand exactly this
situation is predicted in the light gluino scenario.  To see this
we run over the allowed $\tan(\beta)$ range of \ref{eq:tanb} and the
full range of $\mu^2$ allowed by the experimental limits on the
light chargino mass in \ref{eq:mchi}. In addition we require that
the next-to-lightest neutralino, $\chi_1$ have a mass consistent
with limits from the $Z$ width.  The lightest neutralino in this
scenario is the photino, which decouples from the mixing matrix of
the remaining three.  The mass term in the Lagrangian for these
states can be obtained from \cite{reviews} by putting $M_{1/2}$ to
zero. It takes the form
\be
      L = N_3^\dagger {\it M} N_3   \label{eq:L}
\ee
\be
    N_3 = \pmatrix{\tilde{Z}\cr\tilde{H_1^0}\cr\tilde{H_2^0}}
\ee
\be
     {\it M} = M_Z \pmatrix{0&s_\beta&c_\beta\cr
                           s_\beta&0&-\tilde{\mu}\cr
          c_\beta&-\tilde{\mu}&0}      \label{eq:massmatrix}
\ee
where $\tilde{\mu} \equiv \mu / M_Z$ and $s_\beta$ and $c_\beta$
are the sine and cosine of $\beta$ respectively.  The eigenvalues of
${\it M}$ are written in terms of the quantities
\be
      b = \tilde{\mu} \sin{2\beta}     \label{eq:b}
\ee
\be
      a = 1 + \tilde{\mu}^2           \label{eq:a}
\ee
\be
      \cos{\phi} \equiv - { b\over 2} ({a\over 3})^{-3/2}
                                      \label{eq:cosphi}
\ee
The three neutralino eigenstates have masses
\be
 M_n = 2 M_Z \quad\sqrt{a/3} \quad \vert \cos{{\phi + 2\pi n}
 \over 3}\vert \qquad   n=1,2,3             \label{eq:Mn}
\ee
However we relabel the eigenstates according to the definition
$M_1< M_2 < M_3$ so
that $\chi_1$ corresponds to the lightest neutralino apart from the
photino.  This
lightest neutralino is a mixture of the form
\be
    \chi_1 = \alpha_1 \tilde{Z} + \beta_1 \tilde{H_1^0} + \gamma_1
\tilde{H_2^0}         \label{eq:chi1}
\ee
with $\vert\alpha_1\vert^2+\vert\beta_1\vert^2+\vert\gamma_1\vert^2
= 1$. The $\chi_1$ will decay through its $\tilde{Z}$ component into
$q\overline{q}\tilde{g}$ where the $q$ and $q$
should be bottom quarks with the same probability as in
standard model $Z$ decay.  Through its Higgsino components the
$\chi_1$ will decay primarily into $b\overline{b}\tilde{g}$.  Hence
it is possible, if $\alpha_1$ is small enough, that the $b$ excess
at the $Z$ could be due to a production of $\chi_1$ pairs near
threshold.  The experimental mass limits on neutralinos that are
usually quoted \cite{Aleph} do not take into account possible decays
into such states containing a light gluino.  In the light gluino
scenario $\chi_1$ masses near $M_Z/2$ are in fact required in the
$SUGRA$ inspired model.  In Fig.-\ref{fig1} we show the
range of allowed $\mu$ and $M_1$ values with the shape coding
indicating the corresponding value of the lightest chargino mass.
An approximately symmetric set of solutions not shown in
Fig.-\ref{fig2} is found at negative values of $\mu$.  The
points are plotted as squares, triangles, circles, and diamonds if
the lightest chargino is in the 1st, 2nd, 3rd, or 4th quadrant of
the total predicted range $45.5$ $GeV$ to $52.5$ $GeV$.  This very
narrow range overlaps with that required to bring the $R_b$ values
into $1\sigma$ agreement with theory.  (see Fig.-\ref{fig2}
of \cite{WKK}). Such values of the chargino mass will be
definitively tested at $LEP II$ and hence, if the chargino is not
found there, either the light gluino scenario or the supergravity
inspired model for $SUSY$ mass splitting will be ruled out. The
solutions in Fig.-\ref{fig2} with $M_1$ below $45.6$
correspond to on shell production of $\chi_1$ pairs at $LEP I$.  In
Fig.-\ref{fig3}  we show the solution space projected onto
the $\tan(\beta) - \alpha_1^2$ plane with the shape coding
indicating the quadrant values of $M_1$ over its full range of $43$
to $51$ $GeV$. The solutions allowing $Z$ decay into on-shell
$\chi_1$'s are indicated by squares. In the light gluino case
such $\chi_1$ production would appear in the hadronic branching
ratio of the $Z$ and would hence lead to an apparently enhanced
value of the strong coupling constant $\alpha_3(M_Z)$. In a future
paper we will examine quantitatively how much such events might
enhance $R_b$. It is interesting to note from Fig.-\ref{fig3}
that the solutions with low $\alpha_1$ implying an enriched Higgsino
content in $\chi_1$ also have a low $\chi_1$ mass.  The narrow
triangle near $\alpha_1^2 \approx .46$ corresponds to negative
values of the $\mu$ parameter.\hfill\break
\par
     Let us return to the possibility of $R_b$ being related to
light stop quarks and charginos.  In the supergravity related $SUSY$
breaking scheme, the diagonal terms in the sfermion mass matrices
are related to a universal scalar mass $M_0$ and a universal gaugino
mass $M_{1/2}$ by
\be
 M_{\tilde \nu}^2 = M_0^2 + C_{\nu}M_{1/2}^2 +
{1\over 2} M_Z^2 \cos{(2\beta)}   \label{eq:17a}\\
 M_{\tilde{\it l}_L}^2 = M_{\it l}^2 + M_0^2 + C_{\it{l}_L} M_{1/2}^2
 +(-{1\over 2} + \sin{\theta_W}^2 )M_Z^2 \cos{(2\beta)}
 \label{eq:17b}\\
 M_{{\it l}_R}^2 = M_{\it l}^2 + M_0^2 + C_{{\it l}R} M_{1/2}^2 -
\sin{\theta_W}^2 M_Z^2 \cos{(2\beta)}                \label{eq:17c}\\
 M_{{\tilde u}_L}^2 = M_u^2+M_0^2+C_{uL}M_{1/2}^2+\bigl({1\over 2} -
{ 2\over 3} \sin^2{\theta_W}\bigr) M_Z^2 \cos{(2\beta)}
\label{eq:17d}\\
 M_{{\tilde u}_R}^2 = M_u^2 + M_0^2 + C_{uR}M_{1/2}^2 +
{2\over 3} \sin^2{\theta_W} M_Z^2 \cos{(2\beta)}    \label{eq:17e}\\
 M_{{\tilde d}_L}^2 = M_{d}^2 + M_0^2 + C_{dL}M_{1/2}^2 + \bigl(-
{1\over 2} + {1\over 3}\sin^2{\theta_W}\bigr) M_Z^2
\cos{(2\beta)} \label{eq:17f}\\
 M_{{\tilde d}_R}^2 = M_{d}^2 + M_0^2 + C_{dR}M_{1/2}^2 + \bigl(-
{1\over 3}\sin^2{\theta_W} \bigr) M_Z^2 \cos{(2\beta)} \label{eq:17g}
\ee
In addition there are off diagonal terms for each squark of the form
\be
     M_{LR}^2 = A_qM_qM_0      \label{eq:MLR}
\ee
Arguments can be made that the constant $A_q$ should be
$\le 3$ \cite{reviews}.  Then this mixing is negligible except
possibly in the case of the top quark.  One sees that for
sufficiently small $M_0$ and $M_{1/2}$ the stop quark could be
lighter than the top quark.  This occurs naturally in the light
gluino case ($M_1/2\approx 0$) whenever $M_0 \le 500$ $GeV$. The
lightest top quark partner has mass

$$M_{\tilde t}^2 =
M_0^2 + M_t^2 + {1\over 4} M_Z^2 \cos{(2\beta)}
    - {1\over 4} \Bigl [\bigl(M_Z^2 \cos{(2\beta)}(1 -
{8\over 3} \sin^2{\theta_W}) \bigr) ^2 + 16 M_t^2 M_0^2 A_t^2
\Bigr]^{1/2}.$$
\be
\label{eq:mstop}
\ee
Therefore in the monte-carlo described above which runs over all
experimentally allowed values of m and $\tan(\beta)$ we
simultaneously run over values of $M_t$ between $110$ and $200$
$GeV$, over values of $A_t$ between $0$ and $3$ and over values of
$M_0$ between $80$ and $1000$ $GeV$. In view of the $CDF$ results
we  throw out solutions with $M_t<158$ unless $M_{\tilde W}+ M_b
< M_{\tilde t} < M_t$  which would allow the non-standard top decay
mode of \ref{eq:chain} (neglecting the gluino mass).
Fig.-\ref{fig4} shows the correlation between top mass and
lighter stop quark mass for those solutions with $M_{\tilde t}<195$.
Only for low values of stop quark mass is $R_b$ enhanced through
the mechanism $2$ above.  The solution is printed as a square,
triangle, circle, or diamond respectively if $M_0$ is in the $1st,
2nd, 3rd,$ or $4th$ quadrant of the range from $80$ $GeV$ to $550$
$GeV$.  From the work of \cite{BF,WKK} we know that with a lighter
chargino mass about $48$ $GeV$ as predicted here, the maximum
lighter stop quark mass that will reconcile the $R_b$  data is a
steeply falling function of $M_t$. The solutions with a sufficiently
enhanced $R_b$ lie below and to the left of the hatched line in
Fig.-\ref{fig4}. All of these solutions lie in the region
where the top quark has the anomalous decay mode of \ref{eq:chain}
and all of them have a low $SUSY$ threshold ($M_0<400$ $GeV$).\hfill
\break
\par
     In the $M_{1/2}=0$ model, the gluino mass is determined through
radiative corrections in terms of $M_0, A_t, M_t$, and
$M_{\tilde t}$ \cite{CCY}.  In Fig.-\ref{fig5} we show the
correlation between $M_0$ and $M_g$. The solutions with an adequately
enhanced $R_b$ are plotted as $x$'s. Gluino masses up to $2$ $GeV$
are found but only a narrow band below $1.7$ $GeV$ is associated
with an enhanced $R_b$.\hfill\break
\par
            Our conclusions are as follows:\hfill\break
\begin{itemize}
\item{1.} If the enhancement of $R_b$ survives further
experimentation and is due to $SUSY$ then either the standard picture
of soft $SUSY$ breaking through universal scalar and gaugino masses
is wrong or $M_{1/2}\approx 0$ and the gluino lies below $1.7$ $GeV$.
\hfill\break
\par
 In this article we have taken the universal gaugino mass, $M_{1/2}$,
to be strictly zero so that all $SUSY$ breaking originates in the
scalar sector.  If $M_{1/2}$ rises above $1$ $GeV$ the lightest
neutralino acquires significant Higgsino components and the gaugino
masses consistent with $LEP$ limits rise rapidly.  For this reason
the $R_b$ enhancement is not consistent with the standard $SUGRA$
inspired $SUSY$ breaking in the heavy gluino case.  \item{2.} If the
$R_b$ enhancement survives and is due to light gluinos then the
$SUSY$ scale is below $400$ $GeV$, the lightest chargino has a mass
below $52$ $GeV$, and the lightest neutralino (apart from the
photino) has a mass between $43$ and $51$ $GeV$. In this scenario,
the $CDF$ events should not be attributable to top quark decay but
instead to background or to $SUSY$ particle production since the top
will decay through the decay chain \ref{eq:chain}
\item{3.} It is not ruled out that the $R_b$ enhancement could be at
least partially due to on-shell gaugino production with
non-negligible Higgsino components. We have not treated this
quantitatively here and we leave a full combined treatment of the
light gluino mechanisms for $b$ enhancement at the $Z$ to a later
paper.\hfill\break
\end{itemize}
\par
     In the course of this analysis we profited from discussions
with P.W. Coulter and G. Kleppe.  This work was supported in part by
the Department of Energy under grant $DE-FG05-84ER40141$.

\figure{
A typical Feynman diagram leading to enhanced b decay of the $Z$.
Such contributions decouple if $M_{\tilde W}$ and $M_{\tilde t}$ are
large compared to $M_Z$.\label{fig1}}
\figure{
Allowed values of the Higgs mixing parameter, $\mu$, and the second
lightest neutralino mass $M_1$ in the light gluino scenario. Shape
coding indicates the corresponding values of the lighter chargino
mass. (See text.) \label{fig2}}
\figure{
Allowed values of the Higgs vev ration $\tan(\beta)$ and the Zino
fraction in the second lightest neutralino, $\chi_1$, in the light
gluino scenario. Shape coding indicates the value of the $\chi_1$
mass.(See text.)\label{fig3}}
\figure{
The correlation between lighter stop quark mass $M_{\tilde t}$
and the top mass $M_t$ for all possible values of the other
parameters in the light gluino scenario. Solutions shown are all
those with $M_{\tilde t}$ less than $195$ $GeV$ as is required in the
light gluino scenario if the universal scalar mass $M_0$ is less
than $550$ $GeV$. Solutions leading to agreement between theory and
experiment for $R_b$ are those to the left of the hatched curve.
Shape coding indicates the value of the $SUSY$ breaking parameter
$M_0$. (See text.) \label{fig4}}
\figure{
Correlation between the gluino mass $M_{\tilde g}$ and the universal
scalar mass $M_0$ in the $M_{1/2} = 0$ case.  Solutions leading to a
sufficiently enhanced value of $R_b$ are indicated by $x$'s.
\label{fig5}}
\end{document}